\documentclass{ws-p8-50x6-00}

\begin{document}

\newcommand{\bbox}[1]{\mbox{\boldmath$#1$}}

\title{An isobar model for $\eta$ photo- and electroproduction
       on the nucleon}

\author{Wen-Tai Chiang and Shin Nan Yang}
\address{Department of Physics, National Taiwan University,
         Taipei 10617, Taiwan}

\author{L. Tiator and D. Drechsel}
\address{Institut f\"ur Kernphysik, Universit\"at Mainz,
         55099 Mainz, Germany}

\maketitle

\abstracts{An isobar model containing Born terms, vector meson exchange and
nucleon resonances is used to analyze recent $\eta$ photoproduction data
for cross section and beam asymmetry, as well as JLab electroproduction
data. Good overall description is achieved up to $Q^2 = 4.0\;
\mbox{(GeV/c)}^2$. Besides the dominant $S_{11}(1535)$ resonance, we show
that the second $S_{11}$ resonance, $S_{11}(1650)$, is also necessary to be
included in order to extract $S_{11}(1535)$ resonance parameters properly.
In addition, the beam asymmetry data allow us to extract very small
($<0.1\%$) $N^* \rightarrow \eta N$ decay branching ratios of
$D_{13}(1520)$ and $F_{15}(1680)$ resonances because of the overwhelming
$s$-wave dominance. The model is implemented as a part of the MAID
program.}

\section{Introduction} \label{sec:Intro}
Electromagnetic eta production on the nucleon, $\gamma N \rightarrow \eta
N$, provides an alternative tool to study $N^*$ besides $\pi N$ scattering
and pion photoproduction. The $\eta N$ state couples to nucleon resonances
with isospin $I = 1/2$ only. Therefore, this process is cleaner and more
suitable to distinguish certain resonances than other processes, e.g., pion
photoproduction. It provides opportunities to access less studied
resonances and the possible ``missing resonances''.

Eta photoproduction at low energy is dominated by the $S_{11}(1535)$
resonance, which is the only nucleon resonance with a substantial decay to
the $\eta N$ channel. Therefore, $\eta$ photoproduction is an ideal process
to study $S_{11}(1535)$ properties. In contrast, $\pi N$ scattering and
pion photoproduction are always interweaved with the $\eta N$ channel
threshold opening, and often produce inconsistent and/or controversial
results.

Recently, precise experimental data of this process have been measured.
These data include total and differential cross sections for $\gamma N
\rightarrow \eta N$ from TAPS (MAMI/Mainz)~\cite{Krusche:1995nv} and
GRAAL~\cite{Renard:2000iv}, as well as beam asymmetries from
GRAAL~\cite{Ajaka:1998zi} and target asymmetries from ELSA
(Bonn)~\cite{Bock:1998rk}. In addition, there are two recent $\eta$
electroproduction data sets from Jefferson
Lab~\cite{Armstrong:1998wg,Thompson:2000by}.

\section{Isobar Model} \label{sec:Model}
The isobar model used in this work is closely related to the unitary isobar
model (UIM) developed by Drechsel {\it et al.}~\cite{Drechsel:1998hk}. The
major difference is that in the UIM, which deals with pion photo- and
electroproduction, the phases of the multipole amplitudes are adjusted to
the corresponding pion-nucleon elastic scattering phases, while in the
$\eta$ production the unitarization procedure is not feasible since the
eta-nucleon scattering information is not experimentally available.

The nonresonant background contains the usual \emph{Born terms} and
\emph{vector meson exchange} contributions, and can be obtained by
evaluating the Feynman diagrams derived from an effective Lagrangian. In
addition to the dominant $S_{11}(1535)$ nucleon resonance, we also consider
resonance contributions from $D_{13}(1520)$, $S_{11}(1650)$,
$D_{15}(1675)$, $F_{15}(1680)$, $D_{13}(1700)$, $P_{11}(1710)$, and
$P_{13}(1720)$. For the relevant multipoles $A_{\ell\pm}$ ($=E_{\ell\pm},\,
M_{\ell\pm},\, S_{\ell\pm}$) of resonance contributions, we assume a
Breit-Wigner energy dependence of the form
\begin{equation} \label{eq:BWres}
 A_{\ell\pm}(Q^2,W) = \tilde{A}_{\ell\pm}(Q^2)\,
 \frac{\Gamma_{tot}\,W_R}{W_R^2-W^2-iW_R\Gamma_{tot}}\,
 f_{\eta N}(W)\,C_{\eta N}\,,
\end{equation}
where $f_{\eta N}(W)$ is the usual Breit-Wigner factor describing the decay
of the $N^*$ resonance~\cite{Drechsel:1998hk}, and the isospin factor
$C_{\eta N}$ is $-1$. The total width $\Gamma_{tot}$ here is taken as the
sum of $\Gamma_{\eta N}+\Gamma_{\pi N}+\Gamma_{\pi\pi N}$.

\section{Results and Discussion} \label{sec:Result}

\subsection{Photoproduction Results} \label{sec:PhoRes}
\begin{figure}[t]
\begin{minipage}[t]{14pc}
 \centering
 \epsfxsize=60mm
 \epsfbox{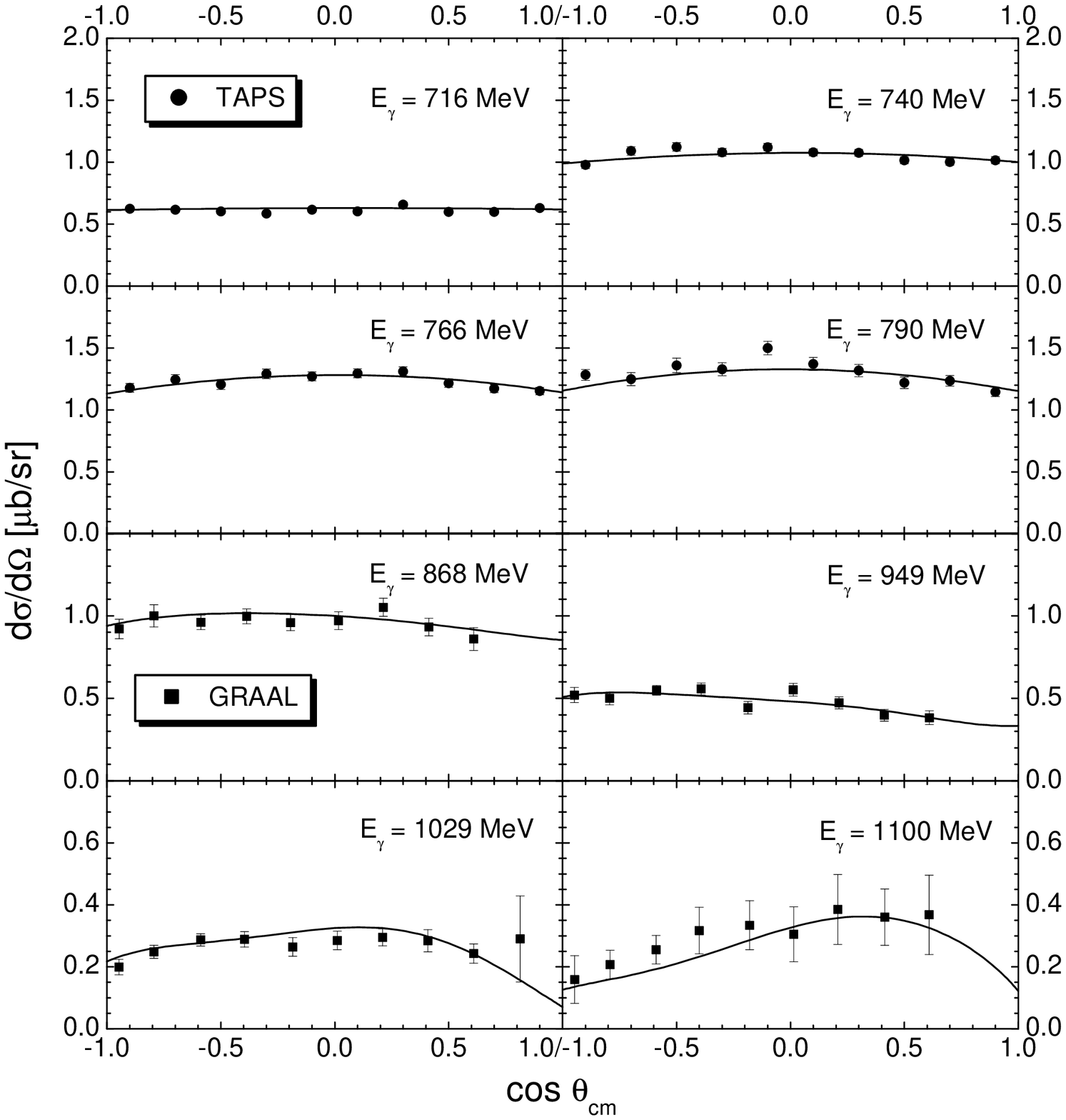}
\end{minipage}
\hspace{\fill}
\begin{minipage}[t]{14pc}
 \centering
 \epsfxsize=55mm
 \epsfbox{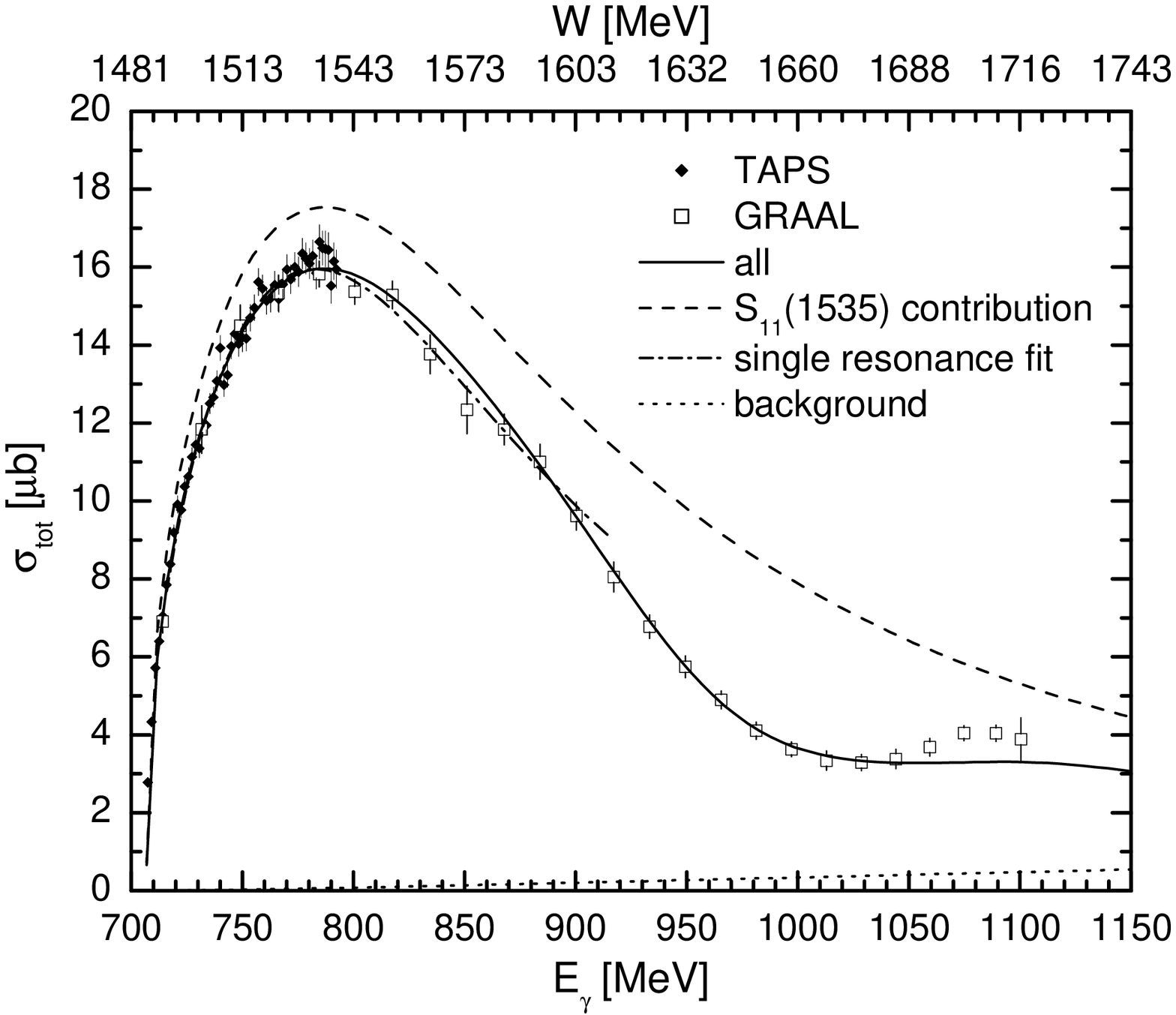}
\end{minipage}
 \caption{Differential and total cross section for $\gamma p\rightarrow
  \eta p$. Data are from TAPS 
  and GRAAL.} 
 \label{fig:cs}
 \vspace{-0.5cm}
\end{figure}

We have fitted recent $\eta$ photoproduction data including total and
differential cross sections from TAPS~\cite{Krusche:1995nv} and
GRAAL~\cite{Renard:2000iv}, as well as the polarized beam asymmetry from
GRAAL~\cite{Ajaka:1998zi} with the isobar model described above. In
Fig.~\ref{fig:cs}, we compare our results of differential cross sections
with the data from TAPS and GRAAL, and they are in very good agreement. In
the low energy region the differential cross section is flat, indicating
the $s$-wave dominance. As the energy goes higher, other partial waves
start to contribute. Note that our result for $E_\gamma^\mathrm{lab}
> 1\ \mbox{GeV}$ shows a sinking behavior at forward angles, which is not
seen in the GRAAL data.

Our result for the total cross section is shown in Fig.~\ref{fig:cs}, and
compared with the TAPS and GRAAL data. Again, they are in good agreement
except that the bump observed from the GRAAL data in the region
$E_\gamma^\mathrm{lab}$ = 1050 - 1100 MeV can not be reproduced from our
model. However, note that the total cross section in the GRAAL data is
obtained from integration of the differential cross sections, using a
polynomial fit in $\cos\theta$ for extrapolation to the uncovered region.
We find that the discrepancy is due to the extrapolation of the GRAAL data
in the forward angles and is not really supported by the data themselves.

In Fig.~\ref{fig:cs}, it is shown that the background contribution is very
small, and the total cross section is dominated by the $S_{11}(1535)$ at
low energy. However, the contribution from the second resonance,
$S_{11}(1650)$, can not be neglected. Even though a single $S_{11}$
resonance can fit the low energy data nicely (the dash-dotted curve in
Fig.~\ref{fig:cs}), it can by no means describe the higher energy region.
Moreover, the single resonance fit yields incorrect resonance parameters.
In fact, the decay width (159 vs. 191 MeV) and photon coupling (103 vs.
$118 \times 10^{-3}\;\mbox{GeV}^{-1/2}$) obtained in the single $S_{11}$
resonance fit are significantly smaller than the full results when both
$S_{11}$ resonances are properly included.

One special feature in polarization measurements of $\eta$ photoproduction
is that through the interference of the dominant $E_{0+}$ multipole with
smaller multipoles, one can access small contributions from particular
resonances.
\begin{figure}[t]
 \centering
 \epsfxsize=65mm
 \epsfbox{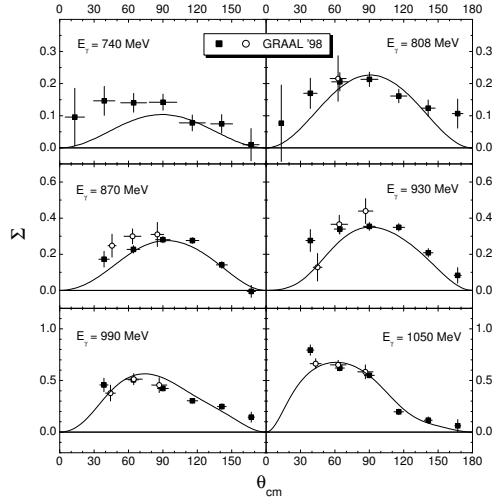}
 \vspace{-0.3cm}
 \caption{Beam asymmetry for $\gamma p\rightarrow \eta p$.
  Data are from GRAAL.}
 \label{fig:beam}
 \vspace{-0.75cm}
\end{figure}
The available beam asymmetry data were measured at
GRAAL~\cite{Ajaka:1998zi} from threshold to $E_\gamma^\mathrm{lab}$ = 1.1
GeV. Higher energy data up to $E_\gamma^\mathrm{lab}$ = 1.5~{GeV} are being
analyzed and will be available soon~\cite{D'Angelo:2001}. In
Fig.~\ref{fig:beam}, we compare our results with these data. An overall
good agreement has been achieved. At low energies, we observe that the beam
asymmetry has a clear $\sin^2\theta$ dependence as a result of interference
between $s$- and $d$-waves. From these low energy data, a branching ratio
of $\beta_{\eta N}=0.06\%$ can be determined for the $D_{13}$(1520). When
energies get higher than $E_\gamma^\mathrm{lab}$ = 930~{MeV}, the data
develop a forward-backward asymmetry behavior, which becomes especially
evident at $E_\gamma^\mathrm{lab}$ = 1050~{MeV}. The $F_{15}$(1680) is
sensitive to this forward-backward asymmetry in $\Sigma$ as discussed by
Tiator {\it et al.}~\cite{Tiator:1999gr}. This is the reason why such a
small branching ratio (0.06\%) can be extracted for this resonance.

\subsection{Electroproduction Results} \label{sec:ElecRes}
\begin{figure}[t]
 \centering
 \epsfxsize=100mm
 \epsfbox{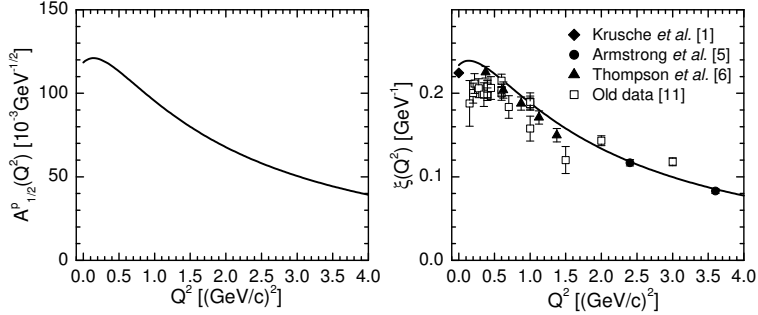}
 \caption{Helicity amplitude $A^p_{1/2}(Q^2)$ for $S_{11}(1535) \rightarrow
  \gamma p$ is shown in the left figure. On the right, we plot the
  quantity $\xi$ ($\equiv \sqrt{\chi\beta_{\eta N}/\Gamma_{tot}}\,
  A_{1/2}^p$), and compare with the extracted values from data.}
 \label{fig:S11Q2}
 \vspace{-0.5cm}
\end{figure}

When fitting recent electroproduction data from
JLab~\cite{Armstrong:1998wg,Thompson:2000by}, we fix all the parameters
determined from the photoproduction data except the $Q^2$ dependence of the
helicity amplitudes $A^p_{1/2,\,3/2}(Q^2)$. The result for the
$S_{11}(1535)$ is shown in Fig.~\ref{fig:S11Q2}.
In order to avoid large model uncertainties arising from different values
of partial and total widths of the $S_{11}(1535)$ employed in other
analyses, we choose not to compare the helicity amplitudes $A^p_{1/2}(Q^2)$
extracted from different analyses. Instead, we compare the
model-independent quantity introduced by Benmerrouche {\it et
al.}~\cite{Benmerrouche:1995uc}, $\xi = \sqrt{\chi\beta_{\eta
N}/\Gamma_{tot}}\, A_{1/2}$, where $\chi = kM/(qM_R)$ is a kinematic
factor. The $\xi$ quantity covers the uncertainty from $\beta_{\eta N}$ and
$\Gamma_{tot}$ between different analyses and is almost independent of the
extraction process. In Fig.~\ref{fig:S11Q2} we compare our $\xi$ values
with the ones extracted from the recent JLab
data~\cite{Armstrong:1998wg,Thompson:2000by} and older data~\cite{Brasse}.
It is seen that overall good agreement is achieved up to $Q^2 =
4.0\;\mbox{(GeV/c)}^2$.
\vspace{-0.25cm}%

\section*{Acknowledgments}
\vspace{-0.10cm}%
W.-T.~C. would like to thank Universit\"at Mainz for the hospitality
extended to him during his visits. This work was supported in parts by the
National Science Council of ROC under Grant No.~NSC89-2112-M002-078, by
Deutsche Forschungsgemeinschaft (SFB 443), and by a joint project NSC/DFG
TAI-113/10/0.

\end{document}